\def\apj{{ApJ}}

\documentclass[cup5b]{caps}
\usepackage{graphicx}
\usepackage{amssymb}
\usepackage{ociwsymp4}
\HeadText{V. V. Smith}  

\begin{document}
\pagenumbering{arabic}

\author[]{VERNE V. SMITH\\Department of Physics, University of Texas
at El Paso}

\chapter{Chemical Evolution in $\omega$ Centauri}

\begin{abstract}

The globular cluster $\omega$ Centauri displays evidence of a complex
star formation history and peculiar internal chemical evolution, setting
it apart from essentially all other globular clusters of the Milky Way.
In this review we discuss the nature of the chemical evolution that
has occurred within $\omega$ Cen and attempt to construct a simple
scenario to explain its chemistry.  We conclude that its chemical
evolutionary history can be understood as that which can occur in a
small galaxy or stellar system ($M\approx 10^{7}\,M_{\odot}$), undergoing
discrete star formation episodes occurring over several Gyr,
with substantial amounts of stellar ejecta lost from the system.   
\end{abstract}

\section{Introduction}

Due to its distinctive chemical evolutionary history, the single globular 
cluster $\omega$ Cen certainly deserves mention
in a meeting on the origin of the elements.  Being the most massive
known Galactic globular cluster, with a mass of $M \approx 3 \times 10^{6}\, 
M_{\odot}$ (Merritt, Meylan, \& Mayor 1997),
it is almost as massive as a small dwarf spheroidal galaxy, such as
Sculptor ($M\approx 6.5 \times 10^{6}\, M_{\odot}$; Mateo 1998).
Indeed, it has been argued that $\omega$ Cen is the surviving remnant
of a larger system, such as a small galaxy, which was captured into
a retrograde Galactic orbit in the distant past (Majewski et al. 2000;
Gnedin et al. 2002).  
It should be mentioned that $\omega$ Cen was the topic of a recent meeting
(in 2001), where all aspects of its
nature were discussed; a large number of topics were covered, not just 
its chemical evolution, and this
broader view of the cluster is well presented in the proceedings
of that meeting (van Leeuwen, Hughes, \& Piotto 2002).

$\omega$ Cen displays a number of fascinating traits that need to be
noted in order to place it within the larger context of chemical
evolution in various types of stellar systems or populations.
First, $\omega$ Cen 
is the only known globular cluster to exhibit a large degree of chemical 
self-enrichment in all elements studied.
Its iron abundance, for example, ranges from
[Fe/H] $\approx$ --2.0, at the low end, up to $\sim$--0.40, 
where the bracket notation
is defined as [A/B] = log($N_{\rm A}$/$N_{\rm B}$)$_{\rm Program Object}$ $-$
log($N_{\rm A}$/$N_{\rm B}$)$_{\rm Sun}$.  The signature of this
abundance spread was first detected as a large width in ($B-V$) color
of the giant branch in $\omega$ Cen from the photographic color-magnitude
diagram of Woolley (1966).  This giant branch ``width'' was
confirmed photoelectrically by Cannon \& Stobie (1973), who suggested
that the color spread was due to varying metallicity.  The inferred
range in the heavy-element abundances (specifically the calcium abundance)
was verified spectroscopically by Freeman \& Rodgers (1975) using
low-resolution spectra. 

A second important trait of $\omega$ Cen is that the abundance
distribution of the elements evolved in a most peculiar way as the
metallicity (e.g., the Fe or Ca abundance) increased.
As noted by Lloyd Evans (1977), the Ba~{\sc ii} $\lambda$4554 \AA\ line
in $\omega$ Cen giants is considerably stronger than in 47 Tuc
giants of similar luminosity and metallicity.  In a following paper,
Lloyd Evans (1983) identified a cool population of giants in
$\omega$ Cen with strong ZrO bands, a characteristic of the $s$-process
heavy-element-rich MS and S stars.  Both Ba and Zr are produced
in slow neutron capture ($s$-process) nucleosynthesis that occurs
during shell He-burning thermal pulses in asymptotic giant branch
(AGB) stars: recent extensive reviews covering the $s$-process and
AGB evolution can be found in Wallerstein et al. (1997) and
Busso, Gallino, \& Wasserburg (1999).   The $\omega$ Cen MS and S stars
identified by Lloyd Evans are of lower luminosity than typical Galactic
or Magellanic Cloud MS and S stars, which are AGB stars undergoing
thermal pulses and the dredge-up of $^{12}$C and $s$-process 
neutron capture elements.  Since the giants found to be $s$-process
rich in $\omega$ Cen are not luminous enough to be thermally pulsing
AGB stars (and, thus, could not have self-enriched their own atmospheres
with $^{12}$C and $s$-process elements), Lloyd Evans (1983) argued that
these chemically peculiar stars, which tend to be the more
metal-rich stars, formed from gas that had been heavily enriched
in $s$-process elements.  Such an enrichment would presumably have been
driven by extensive pollution from a previous population of AGB
stars.

Subsequent high-resolution spectroscopic abundance studies of
$\omega$ Cen giants by Francois, Spite, \& Spite (1988), 
Paltoglou \& Norris (1989), and Vanture, Wallerstein, \& Brown (1994)
found that the $s$-process elemental abundances (such as Y, Zr, Ba,
or Nd) increase as the overall metallicity (i.e. [Fe/H]) increases;
the $s$-process increase is enormous relative to other elements (such as
Ca, Fe, or Ni).  These detailed abundance studies confirm the Lloyd Evans
hypothesis that there is a large $s$-process component involved in the
overall chemical evolution within $\omega$ Cen.

A third trait that defines the character of $\omega$ Cen is that it is
not only chemically peculiar, but also dynamically complex.
Based on Ca abundances derived from low-resolution spectra, Norris,
Freeman, \& Mighell (1996) identified two distinct populations: a
``metal-poor'' component, with [Ca/H] $\approx$ $-$1.4 and containing about
80\% of the stars, and a ``metal-rich'' one comprising the other
20\%, with [Ca/H] $\approx$ $-$0.9.  Later work by Norris et al. (1997),
using radial velocities of the cluster members, revealed that the
metal-poor and metal-rich populations were kinematically distinct.
The metal-poor component is rotating (with $V_{\rm rot}$ $\approx$ 
5 km s$^{-1}$),
while the metal-rich component is not.  The metal-rich population is
also more centrally condensed and has a lower velocity dispersion
(or is kinematically cooler).  Norris et al. (1997) interpreted these
observations as being due to some type of merger in the proto-$\omega$
Cen environment.

More complexity was added to the two-population abundances and kinematics
from Norris et al. (1996, 1997) by the discovery of increasingly
metal-rich components (a third and a fourth) by Pancino et al. (2000),
based on red giant branch (RGB) morphology.  These additional metal-rich
populations comprise about 5\% of all $\omega$ Cen members.  These
newly discovered metal-rich stars became even more interesting when
Ferraro, Bellazzini, \& Pancino (2002) found them to display coherent,
bulk motion with respect to the rest of the cluster stars.  A subpopulation
with its own distinct space motion would suggest a merger, and 
probably a recent one, as playing a role in the evolution of
$\omega$ Cen to its present state.  This potentially fascinating result
has been questioned by Platais et al. (2003), who point out that a
color term has introduced systematic effects in the proper motions
that could be responsible for the apparent bulk motion of the most
metal-rich $\omega$ Cen members.  This question remains open and
no doubt will lead to new observations.

Recent accurate color-magnitude diagrams have also revealed 
details of the star formation history of $\omega$ Cen than can
shed light on the nature of its chemical evolution.  Hughes \&
Wallerstein (2000) used Str\"omgren photometry to derive an
age-metallicity relation.  They found that the metal-poor stars
were typically 3 Gyr older than the metal-rich population, and then
argue that this indicates that $\omega$ Cen enriched itself over this
time scale.  Hilker \& Richtler (2000) also studied a large sample of
$\omega$ Cen stars using Str\"omgren photometry and came to essentially
the same conclusions as Hughes \& Wallerstein (2000), although
Hilker \& Richtler suggest a slightly longer time scale of $\sim$6 Gyr
for chemical enrichment.  A very large color-magnitude diagram study
(with 130,000 stars)
by Lee et al. (2002) points to discrete, multiple stellar populations
in $\omega$ Cen---not necessarily a continuous distribution of ages.
Their modeling of the color-magnitude diagram suggests four
distinct populations in $\omega$ Cen.  This result may then agree
quite nicely with the abundance distributions from Suntzeff \&
Kraft (1996), Norris et al. (1996), or Pancino et al. (2000).
Although Norris et al. (1996) only identified two populations, while
Suntzeff \& Kraft (1996) pointed to a metal-rich tail, the two
most metal-rich components in $\omega$ Cen only comprise a tiny
fraction ($\sim$5\%) of the total number of members.

We will now focus the remainder of this review on trying to understand 
the detailed nature
of chemical evolution in $\omega$ Cen, within the framework of the
metallicities, kinematics, ages, and star formation history as
discussed above.  A simple picture of its 
chemical evolutionary history will be sketched, with the
predictions from this simple model compared to the observed abundances.

\section{The Abundance Distribution and Chemical Enrichment}

Both Suntzeff \& Kraft (1996) and Norris et al. (1996) present relatively
large samples of metallicities in $\omega$ Cen members derived from
low-resolution spectra of the Ca~{\sc ii} infrared (IR) triplet lines.
Both studies endeavored to obtain unbiased samples of stars, with
no selection criteria based on color or abundance.  Norris et al.
(1996) focused on giants brighter than $M_V = -1$ and presented
a sample containing
521 stars.   Suntzeff \& Kraft (1996) studied two separate samples,
one of subgiants with $V$ magnitudes similar to those of the horizontal 
branch (199 members), as well as a sample of bright giants (144 members).
In both studies, the Ca~{\sc ii} equivalent widths were transformed onto
metallicity scales by using calibrating globular clusters (47 Tuc, M4,
NGC 6752 and NGC 6397 for Norris et al. and 47 Tuc, M71, M4, and NGC 6397
for Suntzeff \& Kraft), with Suntzeff \& Kraft (1996) tying their 
scale to the cluster [Fe/H] values, while Norris et al. (1996) opted to
tie their scale to their own calibration in the clusters of [Ca/H].

\begin{figure*}
\centering
\includegraphics[width=1.00\columnwidth,angle=0,clip]{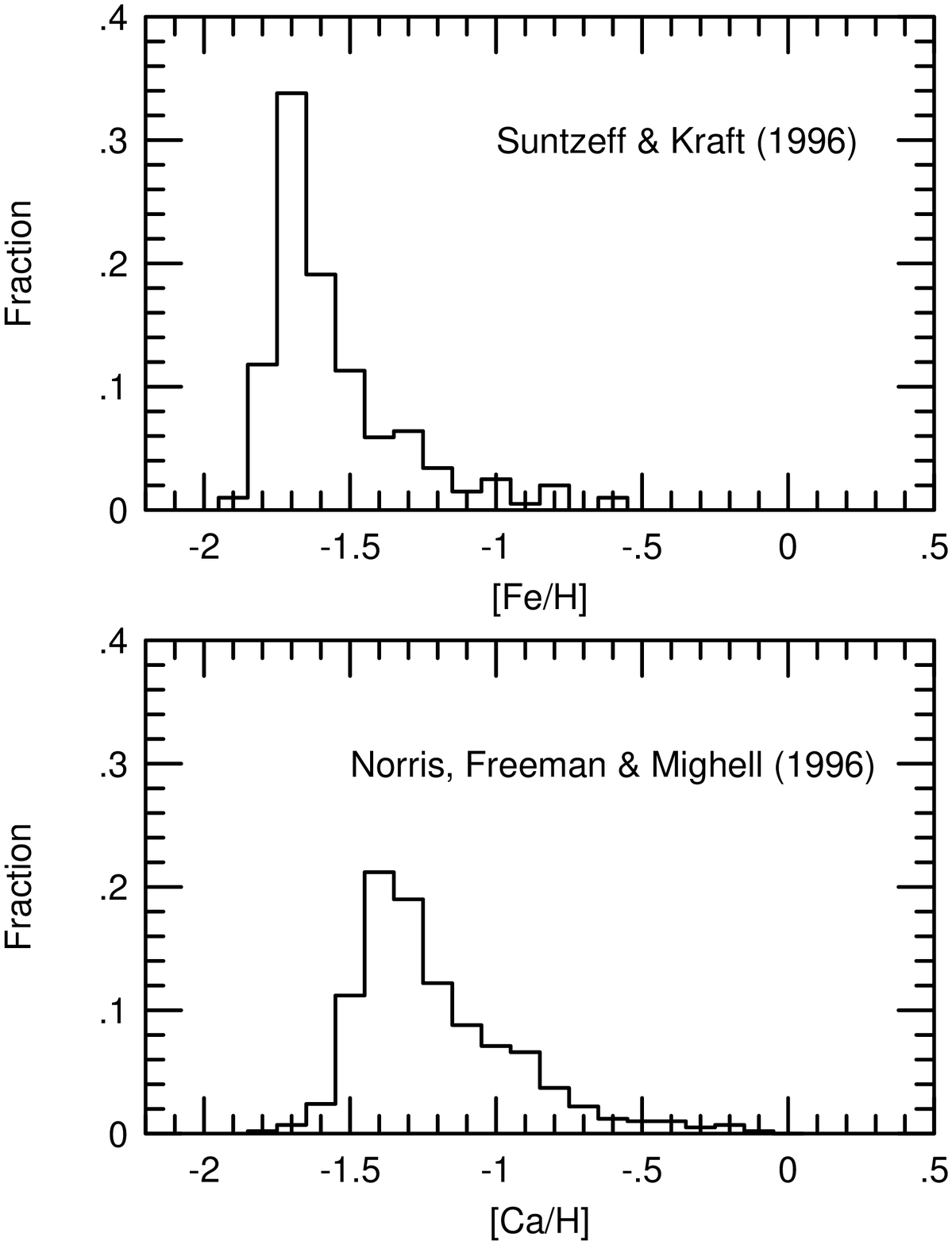}
\vskip 0pt \caption{
Abundance distributions for samples of $\omega$ Cen members
from Suntzeff \& Kraft (1996) and Norris et al. (1996).  The two
``metallicity'' distributions are tied to different scales, with
[Fe/H] for Suntzeff \& Kraft and [Ca/H] for Norris et al., but
their shapes are very similar.  Norris et al. (1996) fit two distinct
populations to their distribution, which basically agrees with the
metal-rich tail noted by Suntzeff \& Kraft (1996). 
}
\end{figure*}

The resulting metallicity distributions from Suntzeff \& Kraft (1996)
and Norris et al. (1996) are shown in Figure 1.1.  Both studies find
essentially the same distribution, but calibrated on different scales
([Fe/H] and [Ca/H]).  There is a sharp cutoff in the number of stars toward
low metallicities, with the sharpness set by the observational
uncertainty.  There is also a well-defined high-metallicity ``tail''
containing about 20\% of the members.  Suntzeff \& Kraft (1996) tried
fitting the metallicity distribution with a simple one-zone model of
chemical evolution with instantaneous recycling.  No satisfactory fit
to the $\omega$ Cen metallicity distribution could be found from such
a model; large effective yields from primary nucleosynthesis products
are needed to fit the metal-rich tail, but such yields result in too
broad of a peak in the metallicity distribution.  Lower effective
yields, which can fit the width of the main metallicity peak, fail to 
fit the metal-rich tail.  Suntzeff \& Kraft do point
out that these two problems can be overcome if two generations of
star formation are considered.  Norris et al. (1996) consider slightly
different types of models, where the cluster enrichment has occurred
within a cloud of gas having some initial heavy-element enrichment,
followed by further enrichment from stellar ejecta, with the total
metallicity distribution being fit by two such components.  Such a
model can fit the [Ca/H] distribution, indicating two rather distinct
populations in $\omega$ Cen.  This population picture 
fits nicely into the later kinematic work from Norris et al. (1997),
which finds two kinematic components that correlate with the 
metallicity, as discussed here in \S 1.1. 

One possible consistent picture for the chemical evolution 
within $\omega$ Cen
that fits the metallicity distributions discussed above would 
contain an initial population of stars with a very narrow range of
metallicity: Suntzeff \& Kraft (1996) estimate $\sigma$[Fe/H]$\le$
0.07 dex, based upon the sharpness of the low-metallicity cutoff.
This narrow metallicity range is similar to what is found in other
globular clusters, where typically $\sigma$[Fe/H]$\le$ 0.05 dex
(Suntzeff 1993).  This initial generation of stars then pollutes
the immediate interstellar medium (ISM) within the proto-$\omega$ Cen, 
and a second generation of stars is formed from this enriched material,
with the new stars forming from different amounts of enriched
material, giving rise to the broad high-metallicity tail.  This
simple picture does not address the cause of the second star formation
episode, or whether such a scenario might be related to some sort of
merger event, but it can, in very broad terms, account for the
overall metallicity distribution.  Additional support for a model
using discrete star formation episodes comes from the more recent
work on the RGB morphology from Pancino et al. (2000) and the
color-magnitude diagram from Lee et al. (2002): both studies
identify four distinct populations.  As the most metal-rich third and fourth
components account for only
$\sim$5\% of the members, just considering a two-epoch star formation 
model, as suggested by Suntzeff \& Kraft (1996) and Norris et al.
(1996), is probably adequate as a start.  Such a model will be considered
by us in \S 1.4; however, before applying this model, a
more detailed discussion of the nature of the various types of
elements involved in the chemical evolution within $\omega$ Cen is
in order. 

\section{Abundance Ratios and the Nature of Chemical Evolution in $\omega$
Cen}

Within a given stellar
population, chemical evolution is driven by nucleosynthesis
averaged over the stellar mass range and subsequent dispersal of
this processed material back into the ISM.    
This heavy-element enrichment over time depends on such processes as 
star formation history, internal stellar evolution and nucleosynthesis
as a function of mass,
how stars return their processed ejecta back into the ISM, and whether
some of the stellar ejecta can be lost from the system. 
By increasing the number of elements considered in an abundance
analysis, we can obtain a more detailed picture of chemical evolution
within $\omega$ Cen.  In particular, elements that arise from
different types of nucleosynthetic processes occurring in stars of
differing masses provide more constraints on the chemical evolution.

In its simplest form, one might consider three basic stellar groups as
contributing to most of a population's chemical evolution:

\begin{itemize}
\item
High-mass stars ($M\ge 8-11\,M_{\odot}$) that explode
as supernovae of Type II (SNe~II) and that contribute much of the
heavy-element enrichment, such as O, Mg, Si, Ca, Ti, and some Fe, as well
as the heavy neutron-rich, rapid neutron capture elements, personified,
for example, by the element europium.
Such stars can contribute their
processed ejecta to a population's chemical enrichment on fairly
short time scales ($\sim 10^{7}-10^{8}$ yr).

\item
Low- and intermediate-mass stars ($M\approx 1.0-8.0\, M_{\odot}$) that evolve onto the RGB and AGB and lose much of
their mass via low-velocity stellar winds as red giants.  Such winds contain
material processed through the stellar interior that has been mixed
to the surface via various red giant dredge-up episodes.  The red giant
winds from AGB stars might contribute significant yields of 
$^{12}$C or $^{14}$N, 
and substantial amounts of the heavy elements produced by
the slow capture of neutrons, the $s$-process, as typified by such
elements as Y, Zr, Ba, or La.  These types of stars will contribute
to chemical evolution over fairly long time scales of $\ge$ 
10$^{8}$--10$^{9}$ yr. 

\item
Supernovae of Type Ia (SNe~Ia), which almost
certainly result from mass transfer in a binary driving a white
dwarf over the Chandrasekhar mass limit.  Such supernovae are
expected to provide very large mass yields of Fe, and these
systems can dominate Fe production in a stellar
population over long time scales of $\sim$1 Gyr. 
\end{itemize}

This is an admittedly thin sketch of stellar nucleosynthesis, but this 
is a limited review and the above points allow us to now investigate
a few of the interesting points concerning chemical evolution within
$\omega$ Cen in light of basic stellar nucleosynthesis.

The ability to detect and analyze a number of different elements, 
some of quite
low abundance or represented by weak spectral lines, requires  
high-resolution spectra, with $R\equiv \lambda/\Delta\lambda \ge 18,000$,
and preferably even higher, with $R=35,000-60,000$.  The results from
high-resolution spectroscopic studies that are discussed here come
from a number of studies that combine both high spectral resolution
and high signal-to-noise ratio (S/N $\approx$ 50--100 or better).  The largest
such study to date is that of Norris \& Da~Costa (1995), who studied
some 40 red giant members, but sacrificed a bit in S/N.  Smaller
samples, but with better S/N, include those of Francois et al. (1988),
Smith, Cunha, \& Lambert (1995), Smith et al. (2000),
or Vanture, Wallerstein, \& Suntzeff (2002).
All of the studies mentioned above include a wide range of elements,
while additional high-resolution analyses by Brown \& Wallerstein (1993)
or Pancino et al. (2002) contain elements produced by SNe~II and SNe~Ia,
but do not include analyses of the $s$-process elements.  Cunha et al. (2002)
probe some 40 $\omega$ Cen giants, but focus on their copper abundances.
All of these high-resolution results will be combined and discussed
in the following two subsections, with the goal being to define the nature
of the nucleosynthesis that is needed to explain the chemical enrichment
observed in $\omega$ Cen. 

\subsection{The $s$-Process and AGB Stars}

\begin{figure*}
\centering
\includegraphics[width=1.00\columnwidth,angle=0,clip]{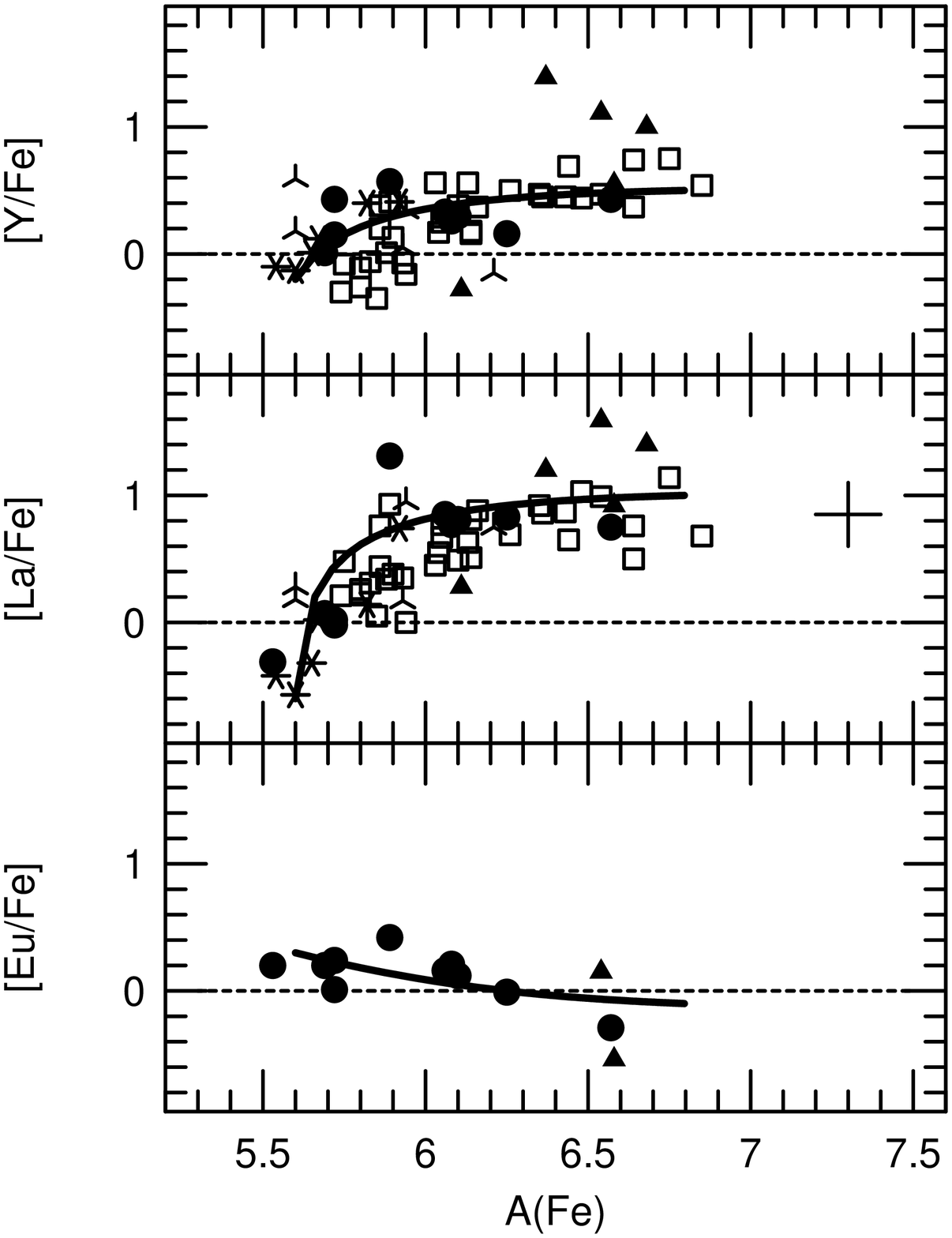}
\vskip 0pt \caption{
Heavy-element abundance ratios in $\omega$ Cen showing 
light and heavy $s$-process elements Y and La, respectively, in the top
and middle panels, along with the $r$-process element Eu in the bottom
panel.  The 6-pointed asterisks are from Francois et al. (1988), open
squares from Norris \& Da Costa (1995), 3-pointed symbols from Smith 
et al. (1995), filled circles from Smith et al. (2000), and filled
triangles from Vanture et al. (2002).  The set of error bars shown in
the middle panel illustrates a typical internal uncertainty. 
As Fe increases by a factor of 10
in abundance, there is an enormous increase in [La/Fe] ($\sim$100 times),
a large increase in [Y/Fe] ($\sim$10 times), and no increase (and perhaps a
slight decrease) in [Eu/Fe].  This is a strong signature of $s$-process
enrichment from low-metallicity, low-mass AGB stars.  The solid curves
in all three panels are what is expected from a chemical evolutionary
scenario discussed in \S 1.4. 
}
\end{figure*}

As first pointed out by Lloyd Evans (1983), the metal-enriched stars
in $\omega$ Cen exhibit an overabundance of $s$-process elements.
This increase is illustrated quantitatively in Figure 1.2, where we
plot [Y/Fe] (top panel), [La/Fe] (middle panel), and [Eu/Fe] (bottom panel) 
versus A(Fe), where A(x) = log~[N(x)/N(H)] + 12, from a number of different
studies.  We note that we have examined each abundance study to
ascertain the sources of their respective $gf$-values and have put
all of these different results on a common scale, when possible (we
also adopt a solar Fe abundance of A(Fe) = 7.50).  Yttrium and lanthanum
are picked to represent the $s$-process, with Y (Z = 39) being a ``light''
$s$-process element and La (Z = 57) being a ``heavy'' $s$-process element.
The abundance ratio of  light to a heavy $s$-process elements is a
diagnostic of the exposure to neutrons experienced by the material. 
Both elements are also spectroscopically well-observed species.
Europium is included as a monitor of $r$-process nucleosynthesis.  It is 
thought that most of the $s$-process abundances result from nucleosynthesis
in AGB stars, while the $r$-process elements are synthesized in SNe~II
(for a review of these neutron capture processes
and their relation to stellar evolution, see Wallerstein et al. 1997).  
Sample error bars are shown in the middle panel of Figure 1.2 to
illustrate approximate internal abundance uncertainties in each study
(adding error bars to each point would overwhelm the plotted points).
All of the high-resolution studies highlighted in these discussions
use similar quality spectra and similar analysis techniques.
As the Fe abundance increases in $\omega$ Cen, the $s$-process component
increases dramatically (Y and La), with no such increase in the
$r$-process (Eu).  The enormous [$s$-process/Fe]
abundance ratios found in the general population of the more metal-rich
$\omega$~Cen stars have not been found in general stellar populations
associated with the Milky Way halo or disk.  Recent abundances
derived in red giants of the Sagitarrius dwarf galaxy by Smecker-Hane
\& McWilliam (2004), however, do show large $s$-process to iron ratios
as observed in $\omega$~Cen.  The solid curves in
Figure 1.2 are predictions from a simple two-component star formation plus
mixing scheme of chemical evolution that is discussed in \S 1.4. 

A key point of Figure 1.2 is that the increase in [La/Fe] is larger than
the increase in [Y/Fe], meaning that the overall ratio of La/Y in
the $s$-process material that has enriched $\omega$ Cen has a value of
[La/Y] $\approx$ +0.4 to +0.5.  This excess of La over Y is a signature
of low-mass, low-metallicity AGB stars driving the $s$-process; Busso
et al. (1999) provide a detailed review of the various types of
$s$-process abundance distributions expected from AGB stars of various
masses and metallicities. 

Smith et al. (2000) compared the heavy-element abundances in
$\omega$ Cen with predictions from
models of AGB nucleosynthesis at the appropriate metallicities
and found that 
1.5 $M_{\odot}$ models yield the best overall fits to the $s$-process
abundance distributions in the $s$-process-enriched members.   
The heavy-element abundances in the
$\omega$~Cen stars thus point to low-mass AGB stars as dominating the
$s$-process enrichment in this cluster.  
The lifetimes of these lower-mass stars
are of order 3 Gyr, pointing to a protracted period of 
star formation,
evolution, and chemical enrichment in $\omega$~Cen.
This time scale agrees with the results from Hughes \& Wallerstein (2000)
and Hilker \& Richtler (2000), who find age spreads of 3--6 Gyr
based upon Str\"omgren photometry and main sequence isochrones (for
Hughes \& Wallerstein), and red giant isochrones (for Hilker \&
Richtler).

\subsection{The $\alpha$-Elements and Contributions from Supernovae}

As reviewed by McWilliam (1997), it has been shown observationally
by many investigations over many years that even-Z elements, such
as O, Mg, Si, S, Ca, and Ti, are overabundant relative to Fe in almost
all Galacic metal-poor stars.  This mix of elements is often referred
to collectively as the $\alpha$-elements, although this collectivization
should not be taken to imply that these elements are all produced in
uniform ratios in a single type of object.  All of the $\alpha$-elements
are produced in SNe~II (e.g., Woosley \& Weaver 1995), but their respective
yields depend on such variables as stellar mass or metallicity.  The
general trend found in the majority of metal-poor Galactic disk and halo
stars is that, for iron abundances of [Fe/H]$\le$ --1.0, the values
of [$\alpha$/Fe] $\approx$ +0.2 to +0.6 and are roughly constant with
metallicity, but having different values of
[$\alpha$/Fe] for the different elements.  At iron abundances 
[Fe/H]$\ge$--1.0, there is a quasi-linear decrease
in [$\alpha$/Fe] toward 0.0 as [Fe/H] approaches 0.0.  The behavior of
[$\alpha$/Fe] versus [Fe/H] is usually interpreted to result from the
time delay between SN~II contributions to chemical evolution relative
to the contributions from SNe~Ia.  The beginning of the decrease in the
values of [$\alpha$/Fe] is then due to the onset of SNe~Ia,
which begin to add large amounts of Fe into a population's pool of heavy
elements. 

Abundances from three elements of the ``$\alpha$ family'' (Si, Ca, and Ti)
in $\omega$ Cen 
are shown in Figure 1.3, plotted as [x/Fe] versus A(Fe).  The top panel
shows [Si/Fe], the middle [Ca/Fe], and the bottom [Ti/Fe] taken from
a number of studies noted in the figure caption.  Missing from this
figure are [O/Fe] and [Mg/Fe] because, as discussed in both Norris \&
Da Costa (1995) and Smith et al. (2000), $\omega$ Cen red giants 
show large ranges in, and anti-correlated behavior between, [O/Fe] and
[Na/Fe], which  is often seen in other globular clusters but not in the
field red giants; for a review of this effect, see Kraft (1994).  This
abundance pattern reveals the effects of H-burning by both the ON part of
the CNO cycles, as well as the Ne-Na cycle.  It is still not clear to
what extent these O and Na (as well as Al and Mg) abundance variations
observed in the globular cluster-like populations may be due to red giant
mixing within the giant itself, or patterns that were imprinted on the
currently observed stars by a previous stellar generation.  Whatever
the fundamental cause, oxygen, sodium, aluminum, and magnesium abundances
in $\omega$ Cen may have been altered by nuclear processes that have
nothing to do with SNe~II; thus, O and Mg are omitted from Figure 1.3. 

In the middle panel are shown horizontal bars that schematically
represent the approximate metallicities [in A(Fe)] of $\omega$ Cen
subpopulations as identified by Pancino et al. (2000).  The most
metal-poor component is labeled RGB-MP, with RGB-Int being the
intermediate-metallicity group and the extremely metal-rich
population (identified by Pancino et al.) is RGB-a.  Recall that
RGB-MP and RGB-Int correspond to the metal-poor and metal-rich
$\omega$ Cen stars identified by Norris et al. (1996) and 
Suntzeff \& Kraft (1996), with these two subpopulations accounting 
for about 95\% of the $\omega$ Cen members.  In the metallicity range
corresponding to RGB-MP and RGB-Int, the $\alpha$-element to iron  
abundance ratios look indistinguishable from those of Galactic halo stars.
Over the range of iron abundances of A(Fe) = 5.5 to 6.5 there are no
strong trends of [Si/Fe], [Ca/Fe], or [Ti/Fe], and the respective 
values of [X/Fe] for these elements in $\omega$ Cen are 
essentially the same as found for the majority of Galactic
metal-poor field halo stars between the same metallicity limits.
There may be slight increases in the values of each of these [$\alpha$/Fe] 
ratios as A(Fe) increases (with slopes of $\sim$+0.1 to +0.2 dex per dex), 
although the significance of these possible trends needs to be assessed with 
a larger,
homogeneous analysis.  The positive values of [$\alpha$/Fe] in the
$\omega$ Cen subpopulations RGB-MP and RGB-Int indicate chemical 
evolution in an environment in which SNe~II control the nucleosynthesis 
(with little, or no significant, contribution from SNe~Ia).  

\begin{figure*}
\centering
\includegraphics[width=1.00\columnwidth,angle=0,clip]{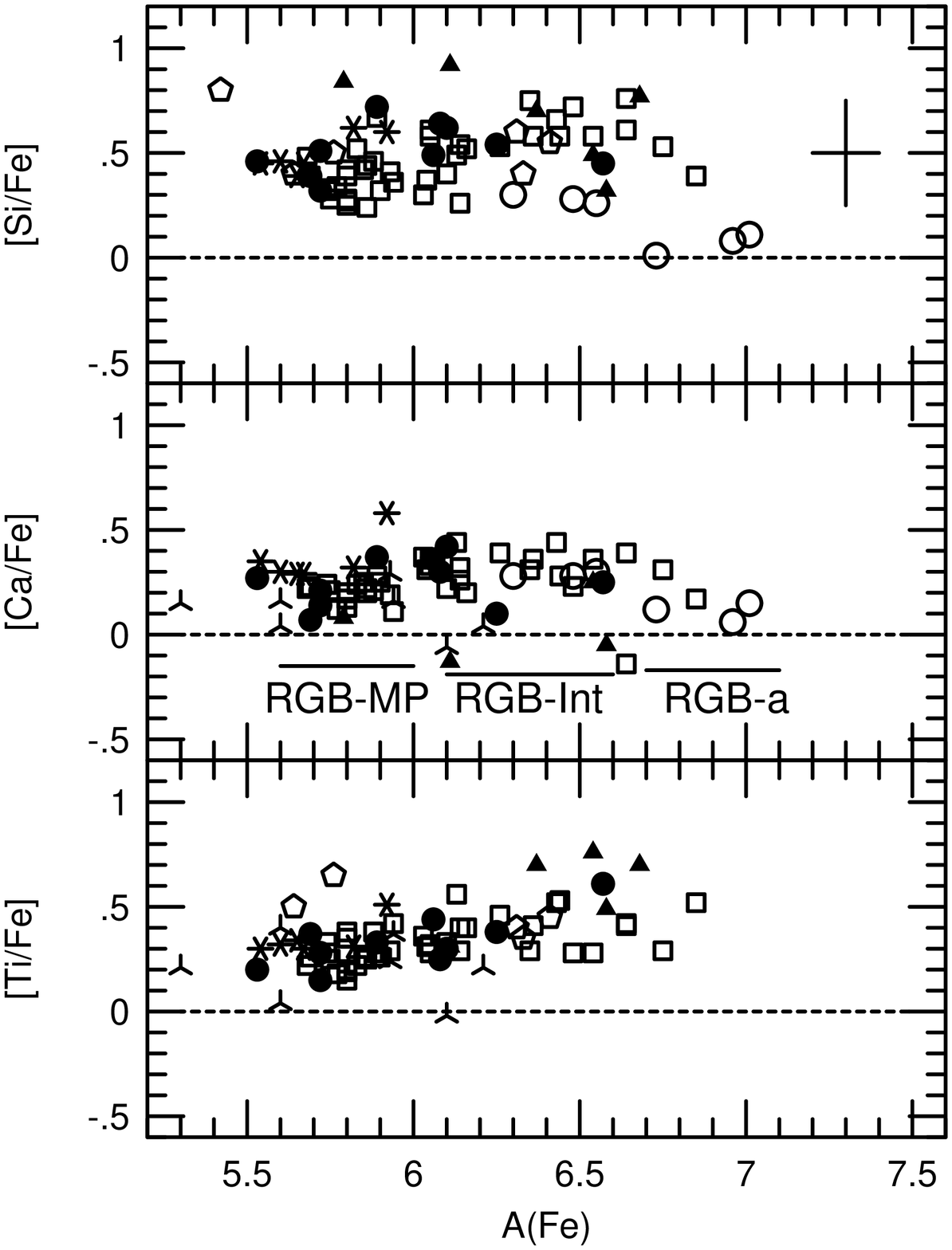}
\vskip 0pt \caption{
Sample elements that are expected to be produced in SNe~II
(Si, Ca, and Ti) compared to Fe.  The 6-pointed stars are from
Francois et al. (1988), open squares Norris \& Da Costa (1995),
open pentagons from Brown \& Wallerstein (1993), 3-pointed symbols
from Smith et al. (1995), filled circles from Smith et al. (2000),
open circles from Pancino et al. (2002), and filled triangles from
Vanture et al. (2002).  A typical internal uncertainty is shown by the
error bars in the top panel.  In the middle panel, the horizontal lines
represent the approximate Fe abundances of the subpopulations identified
by Pancino et al. (2000): RGB-MP (for metal-poor red giants), RGB-Int
(for the intermediate-metallicity giants), and RGB-a (for the more
metal-rich subpopulation).  In the RGB-MP and RGB-Int red giants, all
three of the $\alpha$-elements shown are overabundant relative to Fe,
as found in most Galactic halo stars.  The RGB-a members may show hints
of decreasing values in [Si/Fe] and [Ca/Fe], suggestive of nucleosynthesis
from SNe~Ia.
}
\end{figure*}

The situation for the RGB-a subpopulation may be different, as there
are indications from the results of Pancino et al. (2002), as shown in
Figure 1.3, that [Si/Fe] and [Ca/Fe] may decrease relative to the
values for RGB-MP and RGB-Int.  This may indicate that the RGB-a members
are showing measurable amounts of Fe produced from SNe~Ia.
If so, this is probably not due to a simple time scale difference between
chemical evolution in RGB-MP and RGB-Int compared to RGB-a.  Recall from
\S 1.3.1 that the substantial build-up in the $s$-process abundances
from RGB-MP to RGB-Int requires time scales of $\sim$1--3 Gyr.  This
time scale estimate based on low-mass stellar evolution to the AGB is
also in agreement with the color-magnitude main sequence turn-off
results from Hughes \& Wallerstein (2000) and Hilker \& Richtler (2000). 
Such times are longer than expected for SNe~Ia to affect chemical evolution.
In their study of copper abundance in 40 $\omega$ Cen members, Cunha et al.
(2002) discuss this time scale discrepancy between AGB and SN~Ia
chemical enrichment and speculate on two possible reasons.  One suggestion
is that the stellar density in $\omega$ Cen was so large that binary
systems that would lead eventually to SNe~Ia were disrupted.  Another
possibility is that SN~Ia ejecta have larger kinetic energies than 
SN~II ejecta, leading to lower rates of effective enrichment from SNe~Ia 
compared to SNe~II.  In order to solve this puzzle, a larger and more
homogeneous analysis should be conducted to compare, in detail, the abundance
distributions that characterize the various subpopulations in $\omega$
Cen---especially a study concentrating on the RGB-a members.

\section{A Simple Model for the Chemical Evolution of $\omega$ Cen}

\subsection{$\alpha$-Element and $s$-Process Abundances and Selective 
Retention of Stellar Ejecta}

The increase in the overall Fe abundance in $\omega$ Cen, coupled to
the enhanced ratios of [Si/Fe], [Ca/Fe], or [Ti/Fe]m points to chemical
enrichment from SNe~II.  The possible decrease in both [Si/Fe] and
[Ca/Fe] in the most metal-rich subpopulations of $\omega$ Cen suggests
the eventual appearance of nucleosynthesis products from SNe~Ia.  In the
context of SN enrichment, one puzzle in $\omega$ Cen is how to 
understand the large $s$-process (or AGB) component in its chemical
enrichment.  An initial mass function (IMF) deficient in high-mass stars is certainly one
possibility.  Another hypothesis, perhaps less extreme than altering
the IMF, was explored by Smith et al. (2000) and Cunha et al. (2002).
This hypothesis is that in the lowest-mass systems that undergo
self-enrichment or internal chemical evolution, stellar ejecta or winds
are retained preferentially within the system, or lost to the system,
depending on their ejection velocities.  In this picture, the high-velocity
enriched ejecta from either SNe~II or SNe~Ia are retained inefficiently
within the gravitational potential well of the system when compared to
low-velocity AGB winds (heavily enriched in the $s$-process elements).
Smith et al. (2000) modeled this chemical evolution numerically for a
system with an initial mass of $10^{7}\, M_{\odot}$ in which stars formed 
with the standard Salpeter IMF.  The constraint on mass-ejecta
retention required to fit the $\omega$ Cen abundances was that 10\% of
SN~II ejecta, along with less than 10\% of SN~Ia ejecta (with SNe~Ia
becoming active after a time of 1.2 Gyr), were retained for future
incorporation into new stars.  The $s$-process-rich AGB ejecta were
retained completely, however.  In this simple model, the yields for the
$\alpha$-elements were taken from Woosley \& Weaver (1995) and convolved 
with a Salpeter mass function to determine the chemical compositions of
the SN~II ejecta.  Yields for the $s$-process were taken from the AGB
models discussed in Smith et al. (2000).  This straightforward
exercise reproduces the general trends of SNe (e.g., Si or Ti)
and AGB abundances in $\omega$ Cen, with the only novel assumption
being that SN ejecta are much less efficiently retained than
AGB ejecta.

\subsection{Metallicity-dependent SN II Yields and the Star Formation
History} 

Additional constraints on the nature of chemical evolution in $\omega$ Cen
are provided by elements that show, or are predicted to have, 
metallicity-dependent SN~II yields, such that [element/Fe] values will
depend upon the metallicity of the parent SNe.  Cunha et al.
(2002) studied copper in 40 $\omega$ Cen giants and compared their 
results to those for field stars in the Galactic disk and halo from
Sneden, Gratton, \& Crocker (1991) and Mishenina et al. (2002).  In the
Galactic field stars, [Cu/Fe] shows a steady increase with increasing
[Fe/H] over the range of --2.8 to --1.2 in [Fe/H], with a slope of
$\sim$+0.12 dex per dex in [Cu/Fe] versus [Fe/H].  Sneden et al. (1991)
attribute this general increase in [Cu/Fe] at low [Fe/H] as
due to metallicity-dependent yields for Cu from SNe~II. 
Within the interval
of --1.2 to --1.0 in [Fe/H], [Cu/Fe] increases rapidly from --0.5 to
+0.0, and then stays at this value up to a solar iron abundance.  This 
steep increase in [Cu/Fe] is probably due to Cu production from SNe~Ia.
In contrast to the Galactic field, Cunha et al. (2002) find that [Cu/Fe]
is constant in $\omega$ Cen, at $\sim$--0.55, from [Fe/H] = --2.0 up to
--0.80.  At the higher iron abundances, where [Fe/H] = --1.0 to --0.8,
the $\omega$ Cen stars are falling significantly below the Galactic field
trend (by $\sim$--0.5 dex in [Cu/Fe] at [Fe/H] = --0.8).  Pancino et al.
(2002) also derived [Cu/Fe] in six $\omega$ Cen stars over the range of
--1.2 to --0.4 in [Fe/H].  Their copper abundance values overlap those
of Cunha et al. (2002) nicely and confirm the low values of [Cu/Fe] in
$\omega$ Cen.  The most metal-rich giant in the Pancino et al. sample
is trending upwards in [Cu/Fe] and may signal the measurable addition of
SN~Ia ejecta, as suggested for some of their results for [Si/Fe] and
[Ca/Fe].  Recently, Simmerer et al. (2003) have sampled [Cu/Fe] in
10 ``mono-metallicity'' globular clusters and find that its behavior
with [Fe/H] in these more typical globular clusters is indistinguishable
from the Galactic field.  Simmerer et al. (2003) conclude that the slow
increase in [Cu/Fe] with [Fe/H] at low metallicities results from
metallicity-dependent yields from SNe~II (as indicated by earlier
studies; e.g, Sneden et al. 1991).  In addition, they suggest that the
rapid increase from [Cu/Fe] $\approx$ --0.5 to +0.0 between [Fe/H] = --1.2 to
--1.0 is probably due to the substantial input of Cu from SNe~Ia (as 
argued by Matteucci et al. 1993).

Adding to the interesting differences in [Cu/Fe] between $\omega$ Cen and
the Galactic field stars are the first abundances obtained for fluorine
by Cunha et al. (2003).  These initial results provide tantalizing hints
that fluorine behaves similarly to copper in a comparison of $\omega$ Cen
stars with field-star samples from the Galaxy and the Large Magellanic
Cloud (LMC).  In the case of fluorine, Cunha et al. (2003) found that 
a sample of five $\omega$ Cen giants displayed values of [F/O] that were
significantly lower than the [F/O] ratios found in 
in Galactic and LMC field red giants having
the same oxygen abundances.  They also found that the trend in
[F/O] versus A(O) defined by the Galactic and LMC stars agreed reasonably
well with the predictions of chemical evolutionary models from Timmes,
Woosley, \& Weaver (1995) and Alibes, Labay, \& Canal (2001), in which
fluorine production is driven primarily by neutrino spallation off of
$^{20}$Ne during SN core collapse, as described by Woosley et al. (1990).
These models predict a metallicity-dependent decline in [F/O] versus
A(O) of about --0.30 dex per dex.

Cunha et al. (2003) suggested a scenario to explain the behavior of [F/O] 
in $\omega$ Cen (as well as [Cu/Fe]), in comparison to the field-star
behaviors in the Galaxy and the LMC (for fluorine), by considering the
possible star formation history of $\omega$ Cen.  Their working
assumption is that $\omega$ Cen underwent a few (2--4) well-separated
star formation episodes many Gyr ago: this assumption is motivated by
the morphology of the color-magnitude diagram (e.g., Lee et al. 2002).
In the simplest case of two star formation episodes (that seem to account
for $\sim$95\% of the members, as found in the [Ca/H] distribution from
Norris et al. 1996), some fraction of ejecta from the first stellar
generation is retained with the $\omega$ Cen proto-system (with only a
small fraction of SNe~II retained, even less from SNe~Ia, but 100\% of
AGB ejecta).  This retained ejecta is then mixed with a reservoir of
very metal-poor gas; this reservoir may represent the infall, or merging,
of a primordial (or less chemically evolved) cloud that might itself
induce the second episode of star formation.  If the newly added metal-poor
gas and the ejecta gas are mixed inhomogeneously, stars of the second
generation will have a spread of heavy-element abundances, as observed in
$\omega$ Cen.  Abundance ratios, on the other hand, will be nearly
constant in all stars as long as the infalling (or merging) gas cloud
is severely underabundant in those elements comprising a given ratio.
If an abundance ratio from SNe~II is metallicity dependent, such as [Cu/Fe]
or [F/O], then this ratio will appear anomalous (and will be $\sim$ 
constant with respect to [Fe/H] or [O/H]) when compared to stellar
populations that form from gas that has undergone many star formation
episodes in which stellar ejecta become well mixed with the 
ISM.  In a sense, the abundance ratios carry the ``memory'' of the
metallicities of the SNe~II that formed the bulk of the ejecta out of
which they formed.  A comparison using abundance ratios insensitive to
SN~II metallicity (such as [Si/Fe] or [Ca/Fe]) will appear normal, as
is observed in $\omega$ Cen. 

\subsection{Putting Together the Pieces of the Abundance Puzzle}

The basic picture of an initial population of stars in the proto-$\omega$
Cen environment evolving and depositing ejecta, followed by a second
stellar generation forming from a mixture of this ejecta and added
metal-poor gas, can be tested for consistency by examining abundance
ratios.  In this scenario, the more metal-rich members are stars that formed
from larger fractions of ejecta from the first population.  The limiting
case would be a star that formed from pure first generation ejecta. Thus, 
an approximate abundance distribution for the ejecta (or lower-limit
elemental abundances to this gas) can be taken to be the abundances
defined by the most metal-rich members of the second stellar generation.
The additional metal-poor gas added to this mixture is taken to have 
much lower heavy-element abundances than the ejecta.  The exact composition
of this added gas is unknown, but limits to its metallicity can be
deduced from the sharp low-metallicity cut-offs in the distributions
derived by both Suntzeff \& Kraft (1996) and Norris et al. (1996); see
Figure 1.1.  Both studies found that the sharpness (in either [Fe/H] or
[Ca/H]) of this cut-off is set by their respective observational 
uncertainties: in principle, the low-metallicity cut-off could be a
step function.  This sharp boundary in the metallicity distribution at
low values indicates that the gas in the proto-$\omega$ Cen system was
``pre-enriched'' and well mixed, as discussed in the chemical models
employed by Norris et al. (1996).  It is thus reasonable to use, as the
composition of the added metal-poor gas that is incorporated into the second
generation, this pre-enriched material that was already part of the
proto-$\omega$ Cen system.  In this case, stars that formed from differing
amounts of these two reservoirs of material should follow a ``mixing line''
going from the most metal-poor to the most metal-rich stars.  Referring
back to Figure 1.2, where the heavy-element ratios of [Y/Fe], [La/Fe],
and [Eu/Fe] are shown, the solid curves show such mixing lines, with
the initial abundances taken as those defined by the metal-poor
population.  As discussed in Smith et al. (2000), the heavy-element
abundance distribution in the most metal-poor $\omega$ Cen stars is
characterized by an $r$-process distribution, but these abundances are
then transformed into an $s$-process distribution as [Fe/H] increases.
The $s$-process distribution, with [La/Y] $\approx$ +0.5 and [La/Eu] $\approx$ 1.2,
is indicative of nucleosynthesis from low-mass, low-metallicity
AGB stars, as reviewed by Busso et al. (1999).  The mixing lines shown
in Figure 1.2 are fair representations of the observed abundances of Y, La,
and Eu.

\begin{figure*}
\centering
\includegraphics[width=1.00\columnwidth,angle=0,clip]{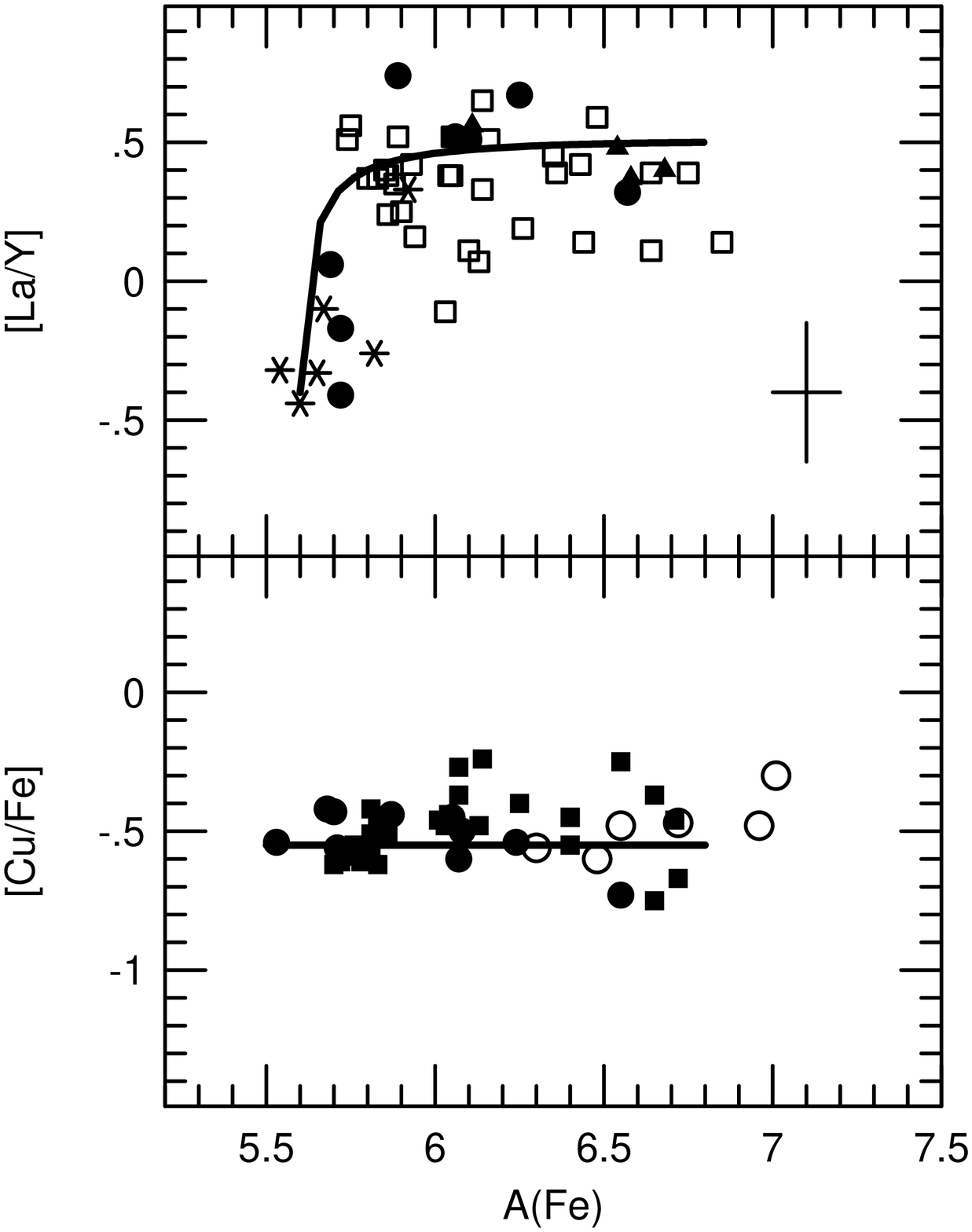}
\vskip 0pt \caption{
The behavior of the heavy-element ratio of [La/Y] versus the
iron abundance (top panel) and that of [Cu/Fe] (bottom panel).
In the top panel, the 6-pointed stars are from Francois et al. (1988),
open squares from Norris \& Da Costa (1995), filled circles from Smith
et al. (2000), and filled triangles from Vanture et al. (2002).  In
the bottom panel, the filled squares are from Cunha et al. (2002) and
open circles from Pancino et al. (2002).  A typical set of error bars
are shown in the top panel.  The increase in [La/Y] as A(Fe) increases
is very steep (in the top panel), illustrating a transition from a
heavy-element distribution dominated by the $r$-process to one dominated
by the $s$-process.  The solid curve is defined by the chemical evolutionary
picture discussed in the text, and this curve tracks the behavior of the
La/Y ratio well.  The [Cu/Fe] values in $\omega$ Cen (bottom panel)
show no measurable change as A(Fe) increases, unlike Galactic halo stars,
which show increasing values of [Cu/Fe] over this metallicity range.  If
copper in these metal-poor populations is dominated by metallicity-dependent
SN~II yields, then the horizontal line is what is expected from the
simple model of chemical evolution invoked here to understand $\omega$
Cen.  In terms of the copper to iron abundance ratios, this model
explains the observations. 
}
\end{figure*} 

Further consistency checks can be conducted using other abundance ratios.
The transition from an $r$-process-dominated heavy-element distribution
in the metal-poor, first stellar generation to a second generation
with a strong $s$-process component should appear as a changing [La/Y]
ratio (the [La/Eu] or [Y/Eu] ratios would be better, but most of the
(few) Eu abundances derived in $\omega$ Cen stars are from the relatively
small sample from Smith et al. 2000).  The top panel of Figure 1.4
illustrates [La/Y] versus A(Fe) from a number of studies of $\omega$ Cen:
note the very rapid rise in [La/Y] as a function of the iron abundance.
The solid curve is the mixing line as set by the same curves used in
Figure 1.2.  Again, the simple two-component mixing model for chemical
evolution in $\omega$ Cen provides an adequate description of the
observed abundances.

The bottom panel of Figure 1.4 shows the results for [Cu/Fe] in $\omega$ Cen
from Cunha et al. (2002) and Pancino et al. (2002).  In Galactic field
halo stars (Sneden et al. 1991; Mishenina et al. 2002) and other
globular cluster stars (Simmerer et al. 2003), values of [Cu/Fe] increase
steadily as A(Fe) increases from $\sim$4.7 to 6.5; this increase is
interpreted as being due to metallicity-dependent SN~II yields (e.g.,
Simmerer et al. 2003).  No such increase is observed in $\omega$ Cen,
and in the picture of a single-metallicity initial stellar population
producing the SN~II products from which a second generation of stars
are born, a constant Cu/Fe ratio would be predicted, with the value of
this ratio set by the metallicity of the first generation of SNe~II.
Such a constant copper to iron ratio is illustrated by the straight
line in Figure 1.4, and this horizontal line is clearly a very good fit to
the behavior of copper with iron.  In addition, the value of [Cu/Fe] $\approx$ 
--0.55 is the value found in field halo stars with A(Fe) $\approx$ 5.5
(see Mishenina et al. 2002), which corresponds to the lowest metallicity
$\omega$ Cen stars.

\subsection{An Overview of Chemical Evolution and Star Formation in
$\omega$ Cen}

In summary, we advocate a picture of chemical evolution in $\omega$ Cen
that is driven by a small number of distinct, time-separated star
formation events; this picture is motivated by the color-magnitude
morphology from, for example, Hughes \& Wallerstein (2000), Hilker \&
Richtler (2000), or Lee et al. (2002).  The large $s$-process abundance
component, relative to elements produced in either SNe~II or SNe~Ia, in
$\omega$ Cen's chemical enrichment points to AGB stars as playing a
major role in chemical evolution (relative to the Galaxy).
This could result from a different IMF, although we prefer to suggest
that selective retention of stellar ejecta is the cause, with SN
ejecta being retained less efficiently than AGB winds.  Continuing
star formation in $\omega$ Cen then results from the mixing of this
stellar ejecta with additional gas arriving from outside of the spatial
regions defined by the stars.  We have taken this gas to be metal-poor,
relative to the stellar ejecta, and find that the observed abundance
trends can be fit by such a picture (the solid curves in Figures 1.2 and
1.4).  We have focused on describing the first two generations in 
$\omega$ Cen for two reasons: (1) these first two stellar generations seem
to account for 95\% of the $\omega$ Cen members (Suntzeff \& Kraft 1996;
Norris et al. 1996; Pancino et al. 2000), and (2) the most metal-rich stars
have yet to be analyzed in detail in terms of their abundance distributions.

The chemical evolution discussed above is similar to a picture suggested
by Freeman (2002).  In this scenario, $\omega$ Cen is the
surviving nucleus of a small galaxy, with its abundance spread being
established by the infall of enriched stellar ejecta from the surrounding
galaxy into the nucleus.  This is a variation from the picture presented 
above in the sense that enriched material is not retained within the
spatial extent of the stars, but arrives from outside.  The Freeman (2002)
scenario may explain more easily the observed kinematic
and angular momentum properties of $\omega$ Cen's subpopulations.
Certainly larger samples of stars and more detailed abundance and
kinematic studies will shed light on the details of the history of
$\omega$ Cen. 

\section{Conclusions}

$\omega$ Cen represents a fascinating and valuable object to those interested
in studying chemical evolution across a variety of environments and
populations.  Although classified as a globular cluster, its complex
star formation and chemical enrichment histories, plus retrograde
Galactic orbit, place it in a different category than the more typical
``mono-metallicity'' globular clusters.  It is almost certainly the
remnant of a captured small galaxy that experienced, before its capture,
a chemical enrichment history unlike that of any other known Galactic
disk or halo population.  Because $\omega$ Cen is relatively nearby,
compared to other small galaxies of the Local Group, it can be used as
a comparison template and laboratory in which to probe different aspects
of chemical evolution that may occur in some small galaxies.

\vspace{0.3cm}
{\bf Acknowledgements}.
VVS acknowledges support for chemical abundance work from the National
Science Foundation (AST99--87374) and NASA (NAG5--9213).

\end{document}